# Optical and microstructural studies of femtosecond laser treated amorphous germanium thin coatings

L. Kotsedi[1,2], A. Abdelmalek[3], V. Bharadwaj[4], C.B. Mtshali[2], Z.Y. Nuru[2,5], B. Sotillo[4,6], G. Coccia[4], S.M. Eaton[4,*], R. Ramponi[7], El-H. Amara[8], M. Maaza[2]

[1]University of the Western Cape, Physics Department, Robert Sobukwe Road, Bellville, South Africa.

[2]iThemba LABS-National Research Foundation, 1 Old Faure road, Somerset-West 7129, P.O. Box 722, Somerset West, Western Cape Province, South Africa.

[3]Physics Department, Theoretical Physics Laboratory, Sciences Faculty, Tlemcen University, Tlemcen 13000, Algeria

[4]CNR - Institute for Photonics and Nanotechnologies (IFN), Piazza Leonardo Da Vinci, 32, 20133, Milano, Italy.

[5]Future Leader- African Independent Researcher, Adigrat University, Department of Physics, P.O. Box 50, Adigrat, Ethiopia.

[6]Departamento de Física de Materiales, Facultad de Ciencias Físicas, Universidad Complutense de Madrid, Ciudad Universitaria, 28040 Madrid, Spain.

[7]Department of Physics, Politecnico di Milano, Piazza Leonardo Da Vinci, 32, 20133 Milano, Italy.

[8]Centre de Développement des Technologies Avancées, CDTA, PO. Box 17 Baba-Hassen, Algiers 16303, Algeria

**\*Corresponding author:** S.M. Eaton (shane.eaton@gmail.com)

**Abstract:**
The study of the relaxation mechanism of amorphous germanium after femtosecond laser irradiation is presented in this work. In particular, a thin germanium coating was deposited onto a glass substrate through the electron beam vacuum coating method. The substrate was kept at room temperature during the coating process, which resulted in a deposited layer characterized by an amorphous microstructure, as observed from the X-ray diffraction. The germanium layer was then irradiated with a femtosecond laser at 1030 nm wavelength, while varying the net fluence from 15 J cm$^{-2}$ to 90 J cm$^{-2}$. Moreover, an extended two temperature model was used to extract the electronic and lattice temperature of the laser heated germanium coating, showing a 32% contribution from heating due to the thermal accumulation effect. The microstructural and morphological studies of the irradiated samples were carried out using θ-2θ X-ray diffraction and high-resolution scanning electron microscopy. From the X-ray diffraction results, it was observed that at higher laser fluence there is an emergence of crystallinity on the germanium layer, with no evidence of oxidation. On the surface, the morphology was observed to evolve to granular sphere, attributed to melting of the material. Finally, an increase in absorbance with laser fluence was observed and the optical band gap of the coating was calculated.

**Keywords**: Germanium; Microstructure; Femtosecond laser; Optical band gap. Two Temperature Model, Lattice temperature.

## 1. Introduction

Understanding the physical and optical response of thin coatings after being exposed to energetic particles (*i.e. charged atoms, electrons, protons or energetic photons*) is important for predicting and tailoring their performance under those conditions. When these energetic particles interact with thin coatings, several physical phenomena can take place depending on the nature of the particles and the type of the coatings. In particular, it has been observed that when energetic photons from a laser source are incident on a thin coating in ambient conditions, the interaction can result in the formation of an oxide layer [1-2], ablation [3-4], or coloration [5-6] depending on the experimental conditions. Optical, crystallographic and microstructural characterization of the laser treated coating can give an insight into the response of different materials to laser processing.

Femtosecond laser processing has been previously applied to modify the surface of thin films [7-8]. The time duration of the pulses used for the fabrication strongly affects the type of modifications experienced by the considered coating, mainly due to a varying strength of the lattice heating [1]. In the same way, the repetition rate of the laser source plays a role on the type of modification produced on a specific coating. This has been observed from the study of the thermal accumulation effect in a material processed with a varying laser repetition rate, resulting in a morphological evolution of the sample due to thermal diffusion acting in concert with the buildup of heat between pulses [9].

Germanium is one of the well-known semiconductors along with Si and GaAs, which recently gained attention due to the possibility to be modified and functionalized through femtosecond laser irradiation [10]. Given the low cost of amorphous germanium compared to a high purity single crystal, the effect of laser irradiation on amorphous germanium (a-Ge) is of great interest, with the goal of creating crystalline germanium (c-Ge) starting from amorphous samples. A previous result from our group [11] has shown an ultrafast change in the optical and thermal properties of a germanium thin film under femtosecond laser irradiation, which was confirmed experimentally by the pump-probe technique [12,13,14]. Crucial for the following calculations is also the demonstration that thermal modification occurs after germanium takes a metallic character upon fs-laser irradiation.

In this study, we present the effect of the interaction of a 1030-nm wavelength femtosecond laser with amorphous germanium coatings. First, the experimental details for the sample preparation and laser treatment will be introduced, together with an extensive characterization of the experimental results. Then, the two-temperature model will be used to calculate the transient temperature of the sample and the excited electron plasma, as well as the thermal accumulation effect under multi-pulse fs-laser irradiation. Eventually, the study of the transformation from the amorphous to the crystalline phase, as well as the deterioration of amorphous germanium under varying net laser fluence will shed light on the potential and viability of this material.

## 2. Experimental and theoretical methods

*2.1. Coating deposition*

Thin germanium coatings with thickness of 2000 Å were grown on a glass substrate. In particular, vacuum coating technology, which uses hot electrons from an electron gun, was used to evaporate a germanium target onto the substrate. Prior to the growth of the coatings, the chamber was pumped down to a base pressure of $1 \times 10^{-7}$ mbar. The germanium target was then de-gassed to remove any atmospheric gases that might have been adsorbed. The substrate temperature during the growth of the coatings was kept at room temperature and the deposition rate of the coatings was measured to be 5 Å/s, while the deposition pressure was maintained at $1 \times 10^{-6}$ mbar.

*2.2. Laser irradiation*

An amplified Yb:KGW femtosecond laser (Pharos, Light Conversion) with 300-fs pulse duration was used to irradiate the thin coatings of germanium. The laser source had a fundamental wavelength of 1030 nm, and the repetition rate was set at 500 kHz. A microscope objective with a 0.2 effective numerical aperture ($NA_{eff}$) was used for focusing the beam. A spot size ($2w_0$) of 42 μm on the sample surface was obtained after the sample was intentionally shifted 105 μm upstream of the focal plane. A 3-axis motorized stage (ABL-1000, Aerotech), which was computer controlled using a CAD software, was employed to execute programmed motion of the sample relative to the incident laser. An area of $0.3 \times 0.3$ cm$^2$ was laser irradiated with linear scans at 20 mm/s separated transversally by 25 μm, to allow sufficient overlap between the adjacent lines. The incident laser power was varied using a computer-controlled attenuator consisting of a half wave plate on a rotation stage followed by a polarizer.

*2.3. Characterization techniques*

The samples were characterized for their optical response, surface morphology, microstructure and depth profile using UV-Vis-NIR absorption, scanning electron microscopy, X-ray diffraction, atomic force microscopy and Rutherford backscattering spectrometry. To study the optical response of the treated sample, CARY 5000 UV-Vis-NIR spectrophotometer was used in an absorbance mode with the scan rate of the measurement set at 600 nm/min, while the light source slit was 50% open. The reference sample was a glass substrate. All the measurements were done at room temperature.

Regarding the morphological study of the modifications, an in-lens detector of the Carl Zeiss Auriga field emission scanning electron microscope was used to generate micrographs from both the backscattered and secondary electrons signal from the interaction of 5 keV accelerated electrons with the sample. The working distance between the detector and the sample was kept at 5.7 mm.

$K_{\alpha 1}$ copper x-rays with a wavelength of 1.54 Å were used to probe the samples' microstructure. This was done using a Bruker D8 advance diffractometer in a θ - 2θ configuration, the scan rate for the measurement was set at 0.1°/min. JCPDS catalogue was used to identify the diffraction peak from the coatings.

Furthermore, coherent and collimated energetic alpha particles accelerated to 3.06 MeV were used to conduct a depth profile study of the laser treated germanium coatings. The backscattered alphas were collected by a detector positioned at 46° from the sample, the beam charge was measured to be approximately 31°C, while the current was 14 nA during the experiment. The samples were mounted on a stepping mechanical motor ladder, which resulted in all the samples measured in sequence without altering experimental conditions.

2.4 Two Temperature Model

The two temperature model (TTM) that tracks the non-equilibrium state of electrons and lattice temperature during ultrashort laser pulse irradiation was employed in this study to analyze the relaxation mechanism of the germanium thin film after laser irradiation. This model is generally used in laser metal interaction, but in this study the model was extended and modified such that it would be fit to use in a semiconducting material like germanium. This was achieved by coupling the standard TTM with the carrier density rate equation in order to take into account the excitation of the valence electrons towards the conduction band during laser irradiation.

The radial and time evolution of excited carrier density $n$ can be defined by the following electron density rate equation, which includes carrier diffusion, photon energy absorption and Auger recombination [15,16]:

$$\frac{\partial n}{\partial t} = \frac{\partial}{\partial z}\left(D \frac{\partial n}{\partial z}\right) + \frac{I(\alpha+\beta I)}{h\upsilon} - \gamma n^3 \qquad (1)$$

where $D$ is the Ambipolar diffusion coefficient, representing the mobility coefficient of charged particles when there is a diffusion driven by an electric field such as the diffusion of electron-hole plasma induced by the laser $E$-field. $\alpha$ is the one-photon absorption coefficient, and $\beta$ is the two-photon absorption coefficient, which in this case can be neglected since the photon energy $h\upsilon$ is greater than Ge's bandgap, $E_g$. $\gamma$ is the Auger recombination coefficient.

To follow the evolution of the energy electronic and lattice subsystem and the thermal behavior of the sample excited under femtosecond laser irradiation, we solve the two following coupled nonlinear equations of the two-temperature model [15,16]:

$$C_e \frac{\partial T_e}{\partial t} = \frac{\partial}{\partial z}\left(k_e \frac{\partial T_e}{\partial z}\right) - G(T_e - T_l) + S \qquad (2)$$

$$C_l \frac{\partial T_l}{\partial t} = \frac{\partial}{\partial z}\left(k_l \frac{\partial T_l}{\partial z}\right) + G(T_e - T_l) \qquad (3)$$

where $S$ is the heat laser source:

$$S = \sqrt{\frac{4ln2}{\pi}}(\alpha + \Theta n)(1-R)\frac{F}{\tau}\exp\left(-(\alpha + \Theta n)z - 4ln2\left(\frac{t}{\tau}\right)^2\right)$$

Note that $e$ and $l$ denote the electron and lattice subsystems, $t$ is the time, $z$ the direction perpendicular to the surface and $F$ is the laser fluence. As previously stated, the sample thickness is $d = 200$ nm. The target is irradiated by $\tau = 300$ fs pulse duration, $\lambda = 1030$ nm wavelength with a repetition rate of $f = 500$ kHz. The thermophysical properties of germanium to be used in the equations are listed in Table A1 (Appendix A). Equation (1), (2) and (3) are solved by the finite different method and implemented by MATLAB software with the following initial and boundary conditions:

$T_e(z, 0) = 300 \text{ K}, T_l(z, 0) = 300 \text{ K}, n(z, 0) = n_0$

$\left.\frac{\partial T_e}{\partial z}\right|_{z=0,d} = \left.\frac{\partial T_l}{\partial z}\right|_{z=0,d} = \left.\frac{\partial n}{\partial z}\right|_{z=0,d} = 0$

## 4. Experimental results

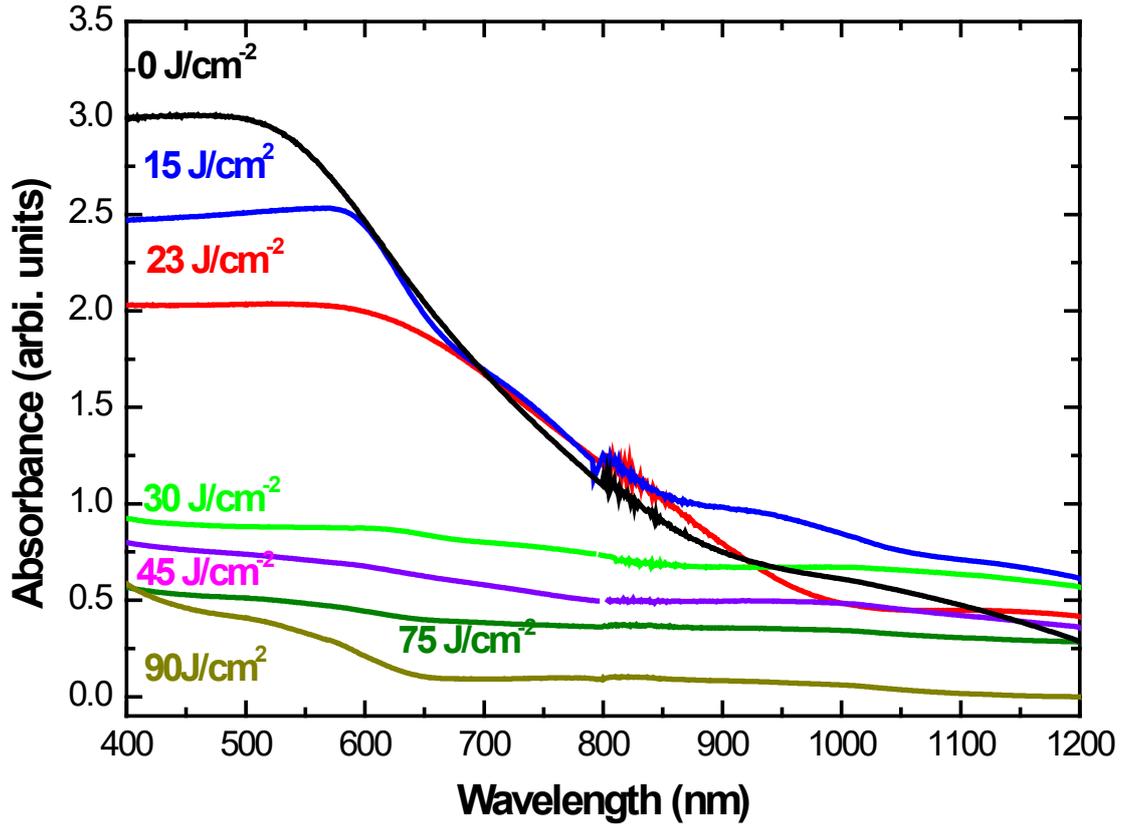

**Fig. 1.** Absorbance spectra of pristine laser irradiated germanium coatings for different laser net-fluences.

*4.1. Optical response*

The absorbance measurements of pristine and laser treated samples are shown in **Fig. 1**. It can be observed that the absorbance of germanium is very low at 1030 nm, but strongly increases at wavelengths lower than 550 nm, close to the second harmonic of the laser fundamental wavelength, consistent with the absorption spectrum of amorphous germanium [9].

From this absorbance measurement it was possible to calculate the optical band gap of the pristine coating, by taking the square of the absorption coefficient versus photon energy through the Tauc plot method. The bandgap value was extrapolated using a straight line to intercept the *x*-axis through the linear gradient of the tail of the plot as shown **Fig. 2**. This resulted in a 1.43 eV band gap for the pristine material, while the optical bandgaps of the laser treated coatings were found to increase with the net fluence. Generally, this increase in optical bandgap has been directly linked to the thickness of the coating [17], as reported in various studies of a-Ge deposited through different deposition techniques [18].

The increase of the optical bandgap, together with the reduction of the absorbance for the irradiated samples with respect to the pristine material would suggest that the laser is mainly ablating the germanium coating, resulting in an increase of the thickness at the edges of the modified areas, together with a reduction of Ge atoms on the coating causing the decrease in absorbance.

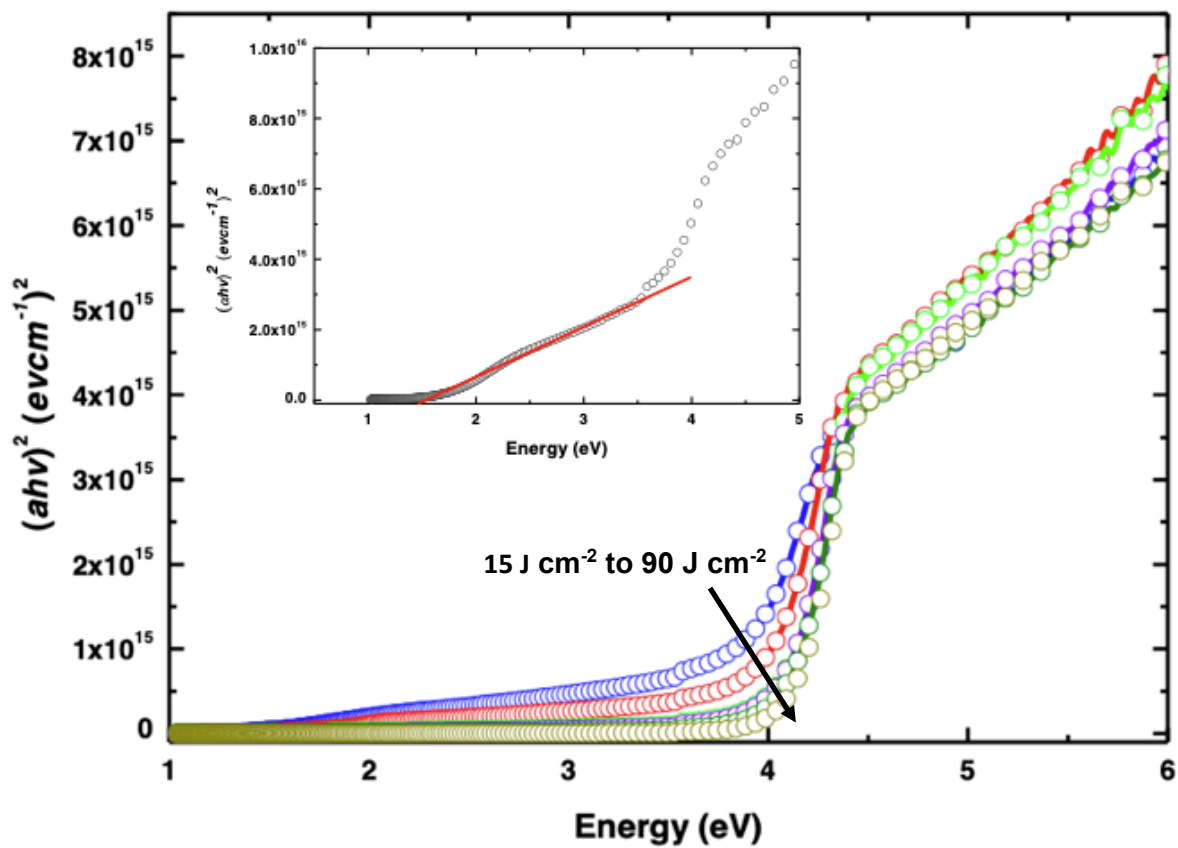

**Fig.2.** Plot of the Tauc optical band gap of the pristine and laser irradiated sample at various fleunces.

Given the interest for the crystallization of amorphous germanium due to laser treatment, the microstructural study of the coatings has been crucial for this study, where X-ray diffraction has been used to characterize the sample modifications. X-ray diffraction patterns of the pristine and laser treated coatings are shown in **Fig. 3**. The diffraction hump centered at $2\theta = 28°$, and the absence of sharp Bragg diffraction peaks indicate the films are amorphous. At higher net fluence of 75 J cm$^{-2}$ and 90 J cm$^{-2}$, there are small Bragg peaks observed at 2θ = 45º and 2θ = 27º, corresponding to (111) and (220) atomic planes. These diffraction peaks are confirmed to be those of c-Ge as reported in the JPDS file number 00-004-0545. Moreover, it can be observed that the Ge crystals are embedded in the amorphous germanium, due to the persistent presence of the a-Ge hump centered at 2θ = 28º. The sizes of the crystallites are smaller, as evidenced by the weak intensity of the Bragg peaks.

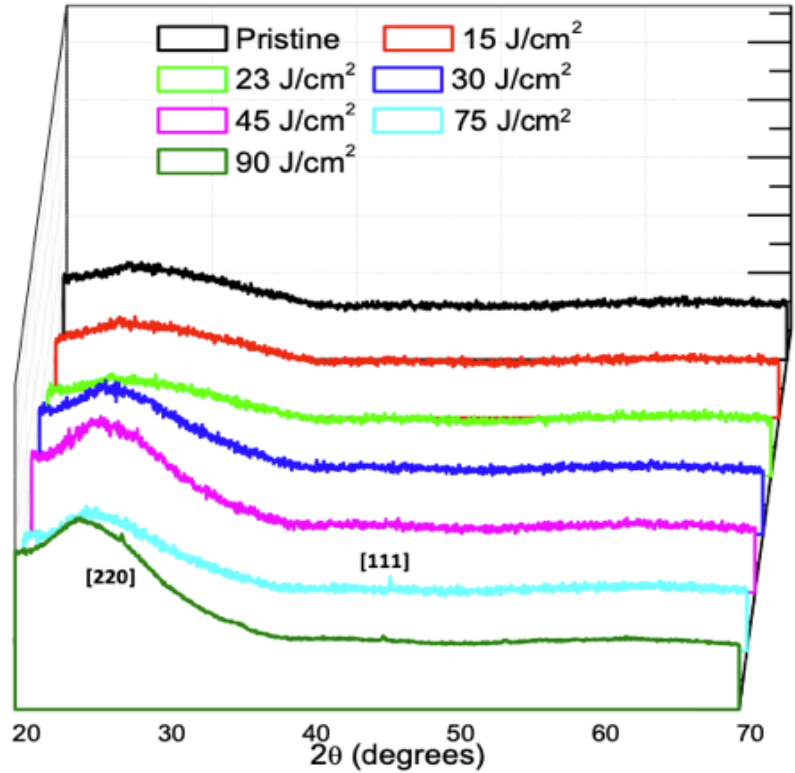

**Fig. 3**. Plot of the X-ray diffraction patterns of pristine and laser irradiated germanium coatings.

This result of a-Ge crystallization is similar to the findings of a previous study where a 800 nm femtosecond laser was used to treat the coatings of a-Si [19] and a crystallization of the amorphous coatings was observed at higher fluences. The Raman spectroscopy of the coatings clearly showed the evolution of the coating from amorphous to crystalline phase, while no formation of an oxide layer at higher laser fluence was observed from the Raman spectroscopy results. This is contrary to the results obtained with other metals like chromium [7] and molybdenum [8], whose coatings resulted in the formation of an oxide layer when exposed to femtosecond laser irradiation. The oxidation in these systems is possible due to high absorbance of these materials. In this case, from the X-ray diffraction patterns of **Fig. 3** by comparing the Bragg peak of the 75 J cm$^{-2}$ and 90 J cm$^{-2}$ laser irradiated samples with the Miller indices for the Bragg peak of germanium oxide from JPDS file number 00-036-1463, it can be confirmed that no layer of germanium oxide was formed due to laser treatment, in agreement with the absorption measurements.

*4.3. Surface morphology*

It has already been shown how the interaction of different laser fluences with the germanium coating produces different types of modifications, making a morphological study of the treated areas necessary. From the SEM images in **Fig. 4**, it can be observed that at a net fluence of 23 J cm$^{-2}$, there is melting of the a-Ge coatings, as can be deduced by the liquid holes on the surface of the coatings. As the net fluence increases, also the heat dissipation area becomes larger, until at 45 J cm$^{-2}$ fluence the liquid reaches the superheated state. In this regime, the ablation results from the thermal accumulation effect, where matter is ejected in the form of vapor and aggregates and then re-deposited around the focal volume, as already deduced from the absorbance measurement. The role of thermal accumulation effect and the calculation of the incubation coefficient will be investigated theoretically in the following sections.

From the SEM images emerges also the presence of wrinkles at the edge of the ridges. These might be due to the formation of LIPSS (laser induced periodic surface structures) during multi-pulse fs laser irradiation, in which a surface plasmon wave can be excited. Recent work confirmed the observation of highly regular nanogratings on the surface of a-Ge under fs laser irradiation, where the coexistence of amorphous and nanocrystalline phases in the LIPSS patterned areas was further confirmed by Raman imaging [20,21]. In the center of the modifications this type of nanostructure is destroyed by the removal of the material and the thermal accumulation effect. However, in the edges of the treated areas these wrinkles can be created due to the weaker laser fluence because of the Gaussian laser profile. This phenomenon has been well demonstrated in our previous work [22], where it was shown that in the LIPSS the direction of the wrinkles is affected by the orientation of the polarization, generally perpendicular to the electric field of the laser beam.

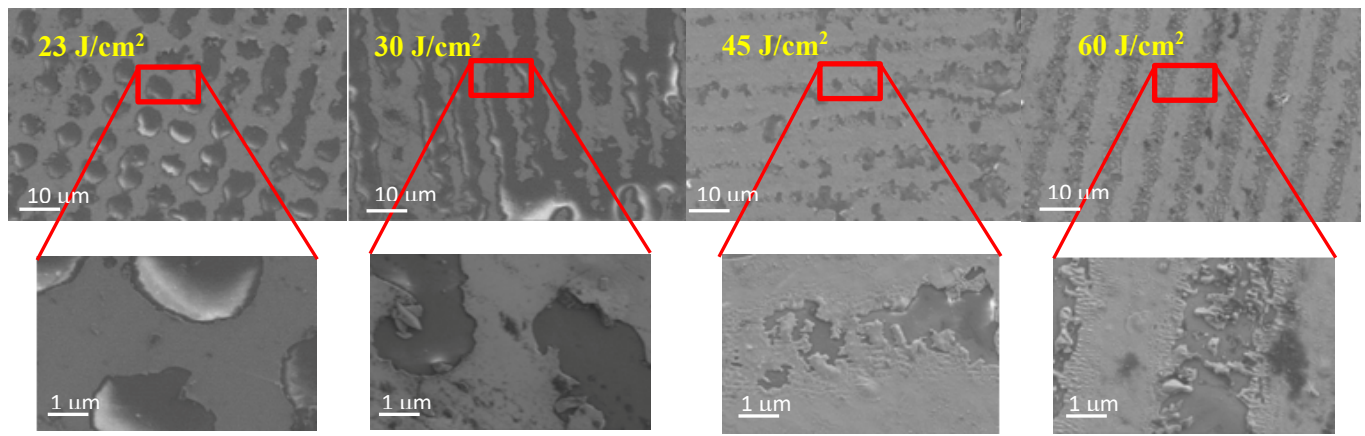

**Fig. 4.** SEM images of laser irradiated coating at different net fluences.

From the Rutherford backscattering spectra shown in **Fig. 5** it can be observed that there is a decrease in the normalized yield of the backscattered germanium atoms as the net fluence increases. This is due to the decrease in the germanium atoms available to backscatter the alpha particles, pointing again to the ablation of the material due to the laser treatment. Furthermore, it can be observed that the energy used for the measurement was 3.06 MeV, which is the oxygen resonance energy. The absence of an oxygen peak for all the samples is further evidence that the coatings were not oxidized during the femtosecond irradiation.

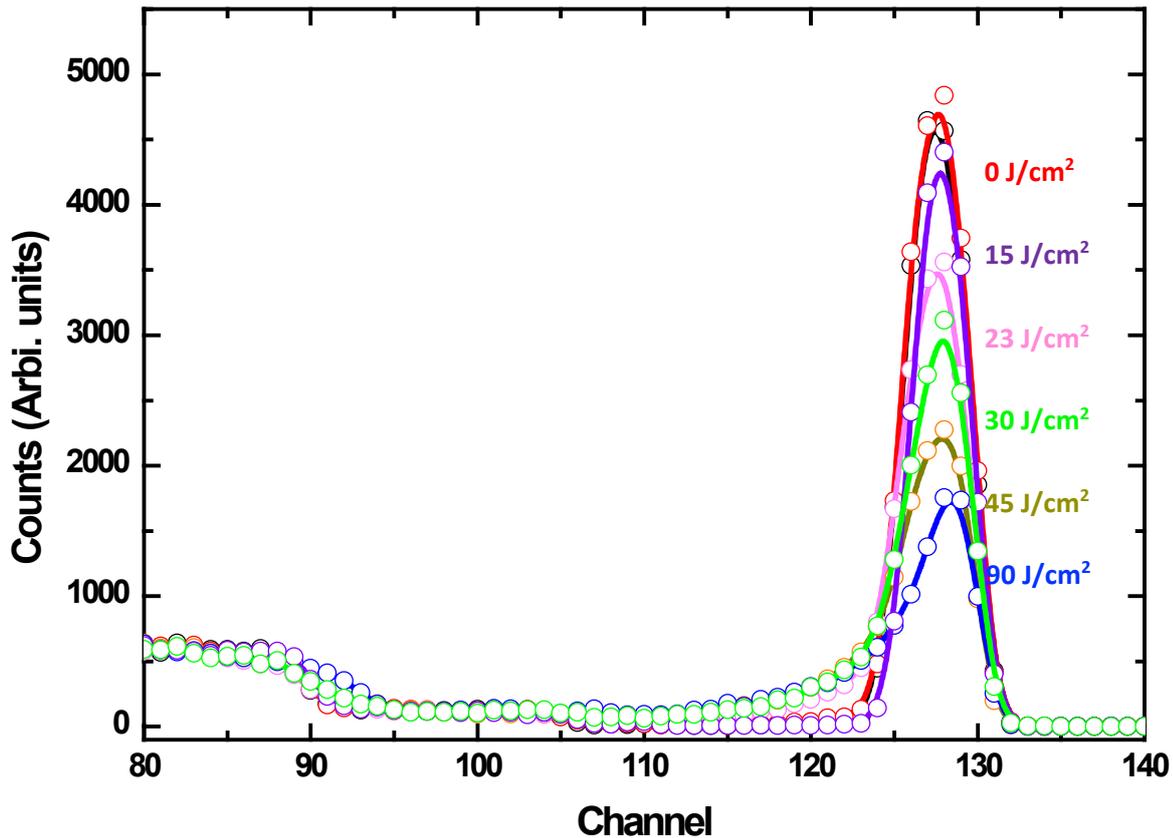

**Fig. 5.** Plot of the Rutherford backscattering spectrometry showing evolution of Germanium after heating with varying laser fluence.

## 3. Theoretical results

It is well known [23] that crystallized solids have well defined edges and faces, diffract X-rays and have defined melting points. In contrast, amorphous solids have irregular or curved surfaces, do not exhibit well-resolved X-ray diffraction patterns and melt over a wide range of temperatures. In summary, amorphous solids are not characterized by well-defined optical and thermal properties, making it a challenging the task of modeling their behavior by TTM differently from the case of crystals. However, we have already confirmed through the X-ray diffraction measurements that the crystalline state can been achieved in amorphous-Ge films under femtosecond laser irradiation (**Fig. 1**) and as also confirmed recently by Bronnikov *et al.* [20]. In particular, high fluences were corresponding to narrow peaks in the X-ray spectrum, indicating a change in the morphology of Ge bulk, consistent with previous works in different materials such as silicon [24], carbon [25] and $Ge_2Sb_2Te_5$ [26]. Given the difficulties in modeling the behavior of amorphous solids and the possibility to reach the crystalline phase for the sample in our experiment, we model the thermophysical behavior of germanium under fs laser irradiation by considering it as a crystal at high laser fluence. The purpose of the simulation is to determine the incubation factor $\xi$, when focal volume is irradiated by $N = 1050$ pulses at 500 kHz, conditions in which the thermal accumulation effect has a significant impact on the ablation threshold. In addition, we compare the different processes achievable using a laser with single pulse irradiation and in burst mode.

*3.1. Single pulse laser irradiation*

There are two kinds of physical mechanisms induced under single pulse fs-laser irradiation. Cold processes, corresponding to sub-picosecond pulse durations and thermal processes for longer pulse durations [27]. Generally, phase explosion is the most dominant mechanism in metals [28], with optical breakdown phenomena occurring in both dielectrics [29] and semiconductors [16]. Moreover, the heat affected zone (HAZ) is minimal in materials irradiated by a femtosecond laser, because the time for absorption is lower than the electron-phonon relaxation time. However, when the interaction is in multipulse mode, the thermal accumulation effect can become significant [30,31], also reducing the ablation precision.

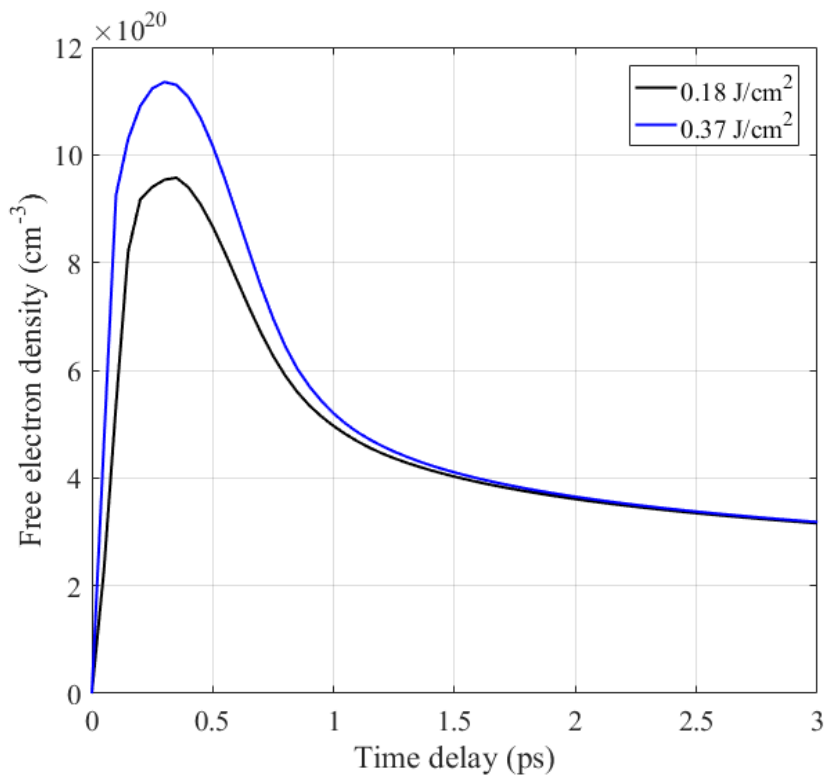

**Fig. 6**: Free electron density evolution of the germanium coating irradiated with single pulse and with different fluences at top center focal volume

**Figure 6** shows the density of free electron plasma excited under single-pulse irradiation with a femtosecond laser. Valence electrons can be excited by direct absorption because the photon energy (1.2 eV) is greater than the c-Ge optical bandgap. The density of free electrons increases considerably during the irradiation and as soon as the pulse is over the density begins to decrease under the Auger recombination effect [32]. It can be noticed how at 0.18 J cm$^{-2}$ the electron density is still lower than the critical density $n_c \approx 10^{21}$ cm$^{-3}$ of metallic state, needing to reach 0.37 J cm$^{-2}$ to exceed the critical value and allow to consider the sample optically as metal-like.

**Figure 7** shows the electron and lattice temperature evolution under single fs-laser pulse irradiation. During the 300 fs-laser irradiation the lattice stays relatively cold, whereas the electron temperature increases up to $10^4$ K. The hot electrons in a cold lattice, induce two non-equilibrium subsystems and therefore will be described by the two-temperature model. This behavior is specific for an ultrashort laser pulse where the interaction with a solid is considered as a non-thermal process. At 0.18 J cm$^{-2}$, the thermal relaxation temperature exceeds the melting temperature ($T_m = 1213$ K). Therefore, it is possible to observe surface damage as a thermal modification around this fluence, as already seen in the experimental part. At 0.37 J cm$^{-2}$, the lattice temperature exceeds the critical temperature ($T_c = 3104$ K), where matter can be ejected by phase explosion and can be considered the ablation threshold according to our model. Table 2 shows a comparison between our TTM result and other previous experimental results, showing good agreement.

**Tab. 2**: The germanium ablation threshold measured in different experiments compared to our theoretical value.

| $\tau$(fs) | $\lambda$(nm) | $F_{th}$ (J cm$^{-2}$) | Ref. |
|---|---|---|---|
| 300 | 1030 | 0.37 | Our TTM |
| 130 | 800 | 0.55 | [12] |
| 200 | 1064 | 0.32 | [39] |
| 150 | 800 | 0.58 | [40] |
| 100-600 | 800 | 0.4 | [41] |

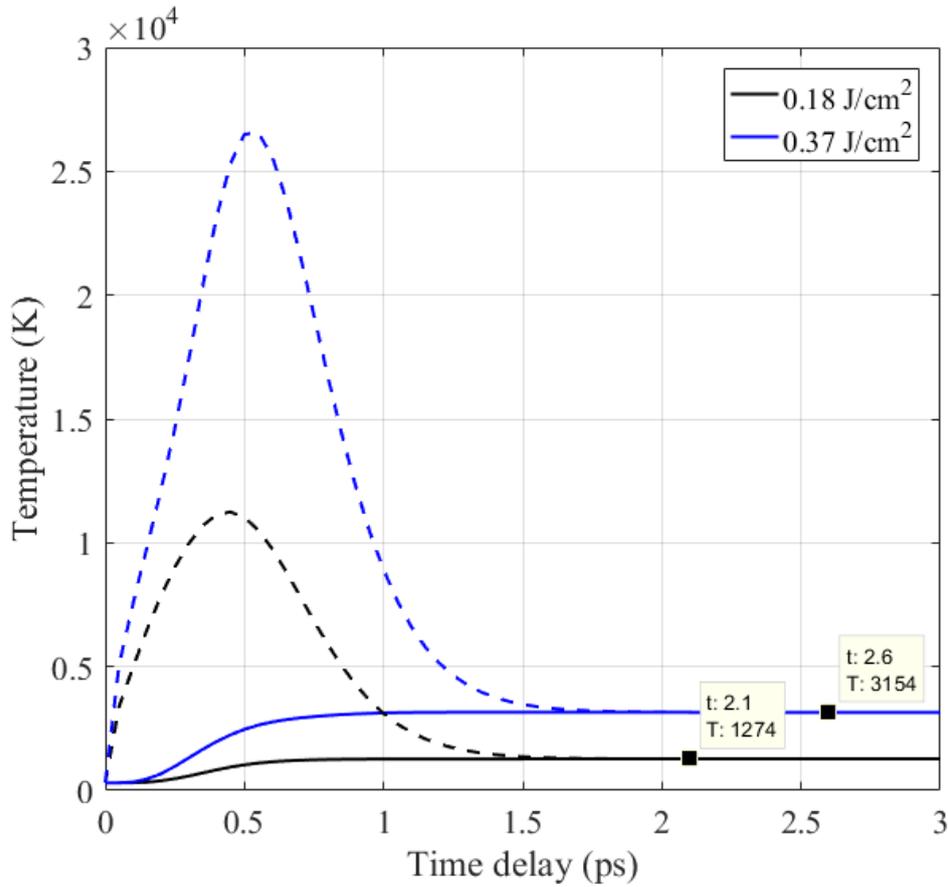

**Fig. 7**: Electron (dashed lines) and lattice (solid lines) temperature evolution under single fs-laser pulse with different fluence.

Recapitulating, when irradiating semiconductors such as germanium by ultrashort laser pulses, electrons in the valence band can absorb energy and be excited to the conduction band via single-photon or multi-photon absorption, depending on the band gap energy [33]. Subsequently, this free electrons-holes plasma mainly recombines via the Auger recombination process, a nonradiative process transferring the excess energy to another free electron-hole pair. As a result, the density of the free electrons progressively decreases, while increasing the kinetic energy of others [32]. At the same time, the free electrons keep transferring their energy to the lattice until thermal equilibrium is reached. This increase in the lattice temperature can induce surface modifications such as melting or formation of bubbles, while the material transitions rapidly to a superheated state or undergoes phase explosions when close to the critical temperature [34]. The phase explosion is the thermal mechanism for material removal when $T_l \geq T_c$ [27], while for electron plasma densities exceeding the critical density $n_c$ [29] laser induced optical breakdown occurs, representing the non-thermal removal mechanism. Thus, as described previously, in Figures 6 and 7 we can observe that during the irradiation, when we get close to the threshold of the laser fluence, the critical density of excited electron plasma is reached, while the material is still cold. This condition allows surface plasmon polariton excitation, leading to a non-thermal modification such as the formation of LIPSS

(Laser Induced Periodic Surface Structures), as shown in [20]. However, we observe that the critical temperature is also reached after thermal relaxation, thus inducing ablation of the material. This results in the formation of aggregates on the irradiated area, as shown in Figure 4. This will inevitably lead to surface damage, which explains why in this situation the nanostructures (LIPSS) are not clearly visible as compared to [45].

*3.2. High repetition rate laser irradiation*

The response of germanium under a single pulse fs-laser irradiation has been presented, the ablation threshold was determined and ultimately the mechanisms responsible for the surface modifications have been described. However, since the repetition rate used in the experiment is quite high (500 kHz), the thermal accumulation effect has a significant impact [30] and should be included in the model. To know the effect of this phenomenon on the ablation threshold, we can calculate the incubation coefficient $\xi$ using the following equation [34], that represents the relation between single-shot threshold fluence $F_{th}(1)$ and *N*-shot threshold fluence $F_{th}(N)$:

$$F_{th}(N) = F_{th}(1)N^{\xi-1} \qquad (4)$$

where $F_{th}(1) = 0.37$ J cm$^{-2}$ is determined by our TTM and $F_{th}(N) = 0.042$ J cm$^{-2}$ corresponding to the net fluence of 45 J cm$^{-2}$ at which the material starts to ablate (see **Fig.4**).

Substituting all the parameters and solving the equation for the incubation coefficient one finds $\xi = 0.68$, where $\xi = 1$ would imply no incubation effect. We can conclude that our sample following the interaction with the femtosecond laser at high repetition rate heats up by 32% due to the thermal accumulation effect. This thermal accumulation effect can increase the heat-affected zone (HAZ) during laser processing and therefore decrease the precision of manufacturing.

*3.3. Burst mode laser irradiation*

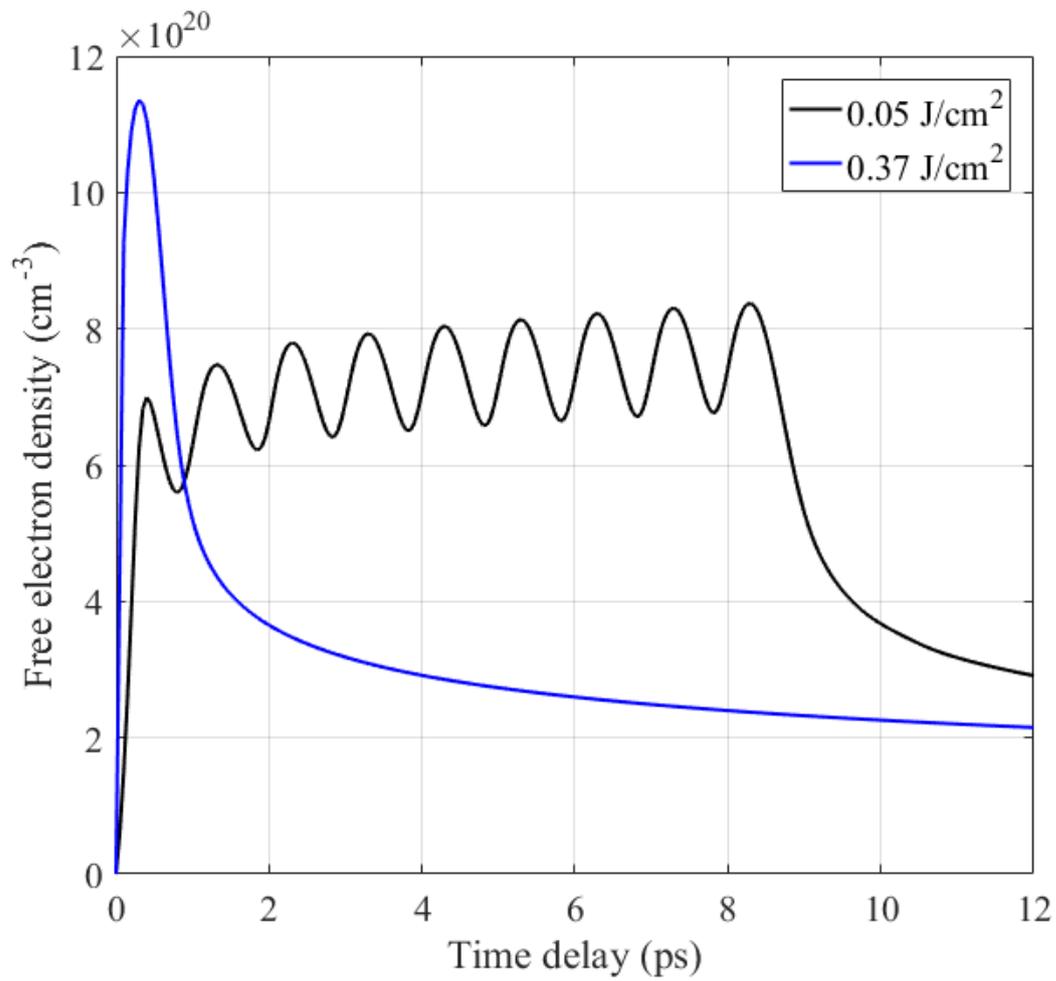

**Fig. 8**. Free electron density time dynamics of germanium sample irradiated under single pulse (0.37 J cm$^{-2}$, blue line) and burst mode (0.05 J cm$^{-2}$, black line).

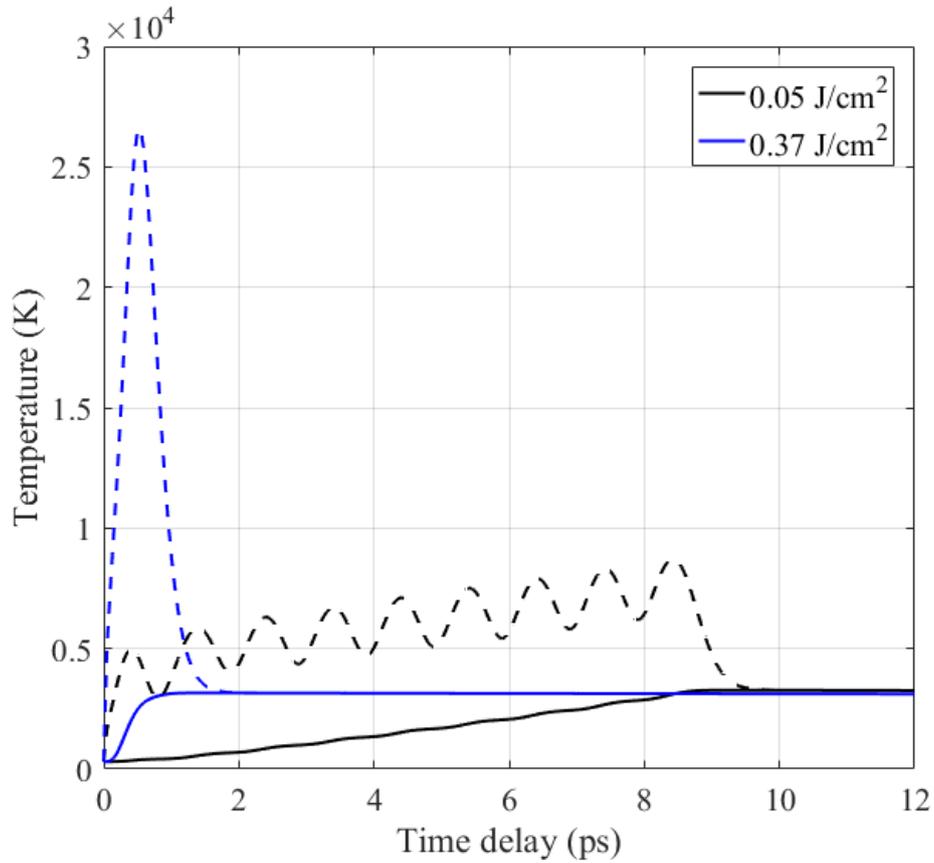

**Fig. 9**. Electron (dashed lines) and lattice (solid lines) temperature evolution under single pulse (0.37 J cm$^{-2}$, blue lines) and burst mode (0.05 J cm$^{-2}$, black lines).

Always more common is the use of ultrafast lasers in burst mode to overcome some obstacles caused by the heat accumulation effect at high repetition rates [36]. A burst pulse train is a set of sub-pulses, all with the same time separation between them, generally in the picosecond time-scale. This technique has shown to reduce the thermal effect with minimal HAZ [37] because the thermal wave does not have enough time to propagate inside the material, as we have also demonstrated in a previous work [38].

**Figures 8** and **9** show respectively the time dynamics of the electron plasma density and the temperature of the material subsystems under single pulse irradiation with 0.37 J cm$^{-2}$ fluence and burst mode with 0.05-J cm$^{-2}$ fluence. We arbitrarily considered nine sub-pulses inside the burst train with a time separation of 1 ps (1 THz repetition frequency). It can be noticed how the plasma density excited by the burst mode does not reach the critical density of the metallic state (**Fig. 8**) and hence the optical breakdown threshold is not reached, differently to what happens for the single pulse irradiation. Moreover, we see that the excited density is moderately stable during irradiation, which indicates that the Auger recombination is nearly constant, giving the free electrons more chance to absorb more laser energy simultaneously.

This alternative mode of irradiation makes it possible to ablate the sample at very low fluence, since the temperature of the lattice reaches the critical temperature at both 0.05 J cm$^{-2}$ in burst mode and 0.37 J cm$^{-2}$ for single pulse (see **Fig. 9**), without reaching the optical breakdown only in the first case. This fast thermal accumulation effect induced during the irradiation in burst mode can be considered as a cold process, because the electronic and lattice sub-systems are not in equilibrium. Therefore, there is no propagation of the thermal wave inside the material, implying a minimum heat-affected zone (HAZ).

Nolte *et al.* [39] propose an experimental study of the incubation effect during the irradiation of a Silicon (Si) surface with fs-bursts at THz frequencies and found out that the ablation threshold can be reduced up to 85% compared to the ordinary pulse mode. Using our modelling data and Eq. (4), we can calculate the incubation coefficient $\xi = 0.089$ for Ge under burst mode, with a reduction of ∼ 91% of the Ge ablation threshold under 9 pulses with burst mode. This result will encourage the scientific community to continue research exploiting this new mode of irradiation to improve the precision of manufacturing materials through ultrafast lasers.

## 4. Conclusions

Thin films of germanium were irradiated using a femtosecond laser with a fundamental wavelength of 1030 nm using various net fluences. The resulting modifications have been characterized through several techniques, giving different insights on the way the laser treatment affected the Ge coating. An interesting result is the change observed in the microstructure of the film through the X-ray diffraction study. In this case, a crystalline phase of germanium emerged at higher net fluences.

Furthermore, the morphological study of the samples showed an evolution of the modifications from small granular spheres due to the heat accumulation to ablation of the coating with increasing net fluence, as observed from the SEM images. Also, the optical response of the considered samples seemed to vary when increasing the net fluence, as shown by the decrease in UV-Vis-NIR absorption. This confirmed on one side the ablation of the coating as the main result of the laser irradiation, while demonstrating the possibility to create germanium crystallites and the absence of Ge oxides after the laser processing.

To gain better insight on the laser irradiation mechanism and analyze what happens switching from continuously pulsing lasers to burst modes ones, the two-temperature model was used to calculate the electronic and lattice temperature of the sample during laser heating, as well as the incubation coefficient. From these calculations, the mechanism of heat transfer was inferred to have a 32% contribution from the heat accumulation effect for the high repetition rates used in the experiment and the surface morphology could be related to heating propagation from the photons to the lattice. Finally, it was demonstrated how the use of ultrafast lasers in burst mode would reduce drastically the incubation coefficient, and hence the ablation threshold, enabling more precise processing of the materials. Indeed, our findings will motivate further research towards the crystallization of a-Ge and burst-mode processing of materials for manufacturing applications.

# Appendix A

**Tab. 1A**: Ge properties used in our modeling from Ref. [10,11,15,43-45].

| Property | Value |
|---|---|
| Ambipolar diffusion coefficient $D$ (m$^2$.s$^{-1}$) | $65 \times 10^{-4}(T_l/300)^{-3/2}$ |
| One-photon absorption coefficient $\alpha$ (m$^{-1}$) | $1.4 \times 10^6 (1 + T_l/2000)$ |
| Bandgap $E_g$ (eV) | $0.743 - \dfrac{0.456 \cdot 10^{-3} T_l^2}{210 + T_l}$ |
| Auger recombination coefficient $\gamma$ (cm$^6$.s$^{-1}$) | $2 \times 10^{-31}$ |
| Heat capacity of electron $C_e$ (J.m$^{-3}$.K$^{-1}$) | $3nk_b$ |
| Drude relaxation time $\tau_D$ (s) | $4 \times 10^{-13}(1 + (n/10^{27})^2)$ |
| Electron-phonon coupling $G$ (J.m$^{-3}$.K$^{-1}$.s$^{-1}$) | $C_e/\tau_D$ |
| Density $\rho$ (kg.m$^{-3}$) | $\begin{cases} -0.1085 T_l + 5409 & T_l < T_m \\ -0.4529 T_l + 6124 & T_l \geq T_m \end{cases}$ |
| Heat capacity of lattice $C_l$ (J.kg$^{-1}$.K$^{-1}$) | $\begin{cases} \dfrac{(1.256 \times 10^6 + 900 T_l - 0.644 T_l^2 + 1.61 \times 10^{-4} T_l^3)}{\rho} & T_l < T_m \\ 3.194 \times 10^{-8} T_l^3 - 1.287 \times 10^{-4} T_l^2 + 0.1774 T_l + 259.4 & T_m \leq T_l < 1500 \text{ K} \\ 343.7225 & T_l > 1500 \text{ K} \end{cases}$ |
| Electron thermal conductivity $k_e$ (W.m$^{-1}$.K$^{-1}$) | $28 e^{-2936/T_e}$ |
| Lattice thermal conductivity $k_l$ (W.m$^{-1}$.K$^{-1}$) | $\begin{cases} 1.027 \times 10^5 T_l^{-1.297} & T_l < T_m \\ -9.458 \times 10^{13} T_l^{-4.263} + 48.56 & T_m \leq T_l < 1473 \text{ K} \end{cases}$ |
| Free-carrier absorption cross-section $\Theta$ (m$^2$) | $6.6 \times 10^{-24}$ |
| Surface reflectivity $R$ (%) | $\begin{cases} 6 \times 10^{-3} T_l + 38048 & T_l < 633.15 \text{ K} \\ 4.57 \times 10^{-3} T_l + 39.375 & 633.15 \leq T_l < T_m \\ 79.2 & T_l > T_m \end{cases}$ |
| Melting temperature $T_m$ (K) | 1211 |
| Critical temperature $T_c$ (K) | 3104 |
| Initial free electron density $n_0$ (cm$^{-3}$) | $2.33 \times 10^{13}$ |

# 5. Acknowledgments

The authors are grateful for support from the National Research Foundation (Republic of South Africa) under the grant number 91470 and bilateral Italy-South Africa project "Novel High Temperature Zr based nano-structured coatings for selective solar absorbers for Concentration Solar Power applications" (code ZA18MO07). S. M. Eaton of CNR-IFN is thankful for financial support from the European Union through the H2020 Marie Curie ITN project LasIonDef (GA No. 956387) and MUR (Ministero dell'Università e della Ricerca) through the projects FISR2019-05178 and PNRR PE0000023 NQSTI. S. M. Eaton is grateful to the Department of Physics at Politecnico di Milano for access to the FELICE laboratory for crucial characterization experiments. V. Bharadwaj is grateful for financial support from the ERC project PAIDEIA GA n.816313. We thank Dr. Luigino Criante and Prof. Guglielmo Lanzani